\documentstyle[12pt]{article}
\catcode`\@=11

\@addtoreset{equation}{section}
\def\theequation{\arabic{section}.\arabic{equation}}

\catcode`\@=11

\def\section{\@startsection{section}{1}{\z@}{3.5ex plus 1ex minus
   .2ex}{2.3ex plus .2ex}{\large\bf}}

\def\eqnarray{\let\@currentlabel=\theequation\refstepcounter{equation}
    \global\@eqnswtrue
    \global\@eqcnt\z@\tabskip\@centering\let\\=\@eqncr
    $$\halign to \displaywidth\bgroup\@eqnsel\hskip\@centering
      $\displaystyle\tabskip\z@{##}$&\global\@eqcnt\@ne
       \hfil${{}##{}}$\hfil
      &\global\@eqcnt\tw@ $\displaystyle\tabskip\z@{##}$\hfil
       \tabskip\@centering&\llap{##}\tabskip\z@\cr}
\def\lefteqn#1{\hbox to 4\arraycolsep{$\displaystyle #1$\hss}}
\def\thesection{\arabic{section}.}

\def\appendix{\setcounter{section}{0}
        \def\thesection{Appendix.}
        \def\theequation{\Alph{section}.\arabic{equation}}}

\long\def\@makefntext#1{\parindent 0cm\noindent
\hbox to 1em{\hss$^{\@thefnmark}$}#1}
\def\IR{{\hbox{{\rm I}\kern-.2em\hbox{\rm R}}}}
\def\IH{{\hbox{{\rm I}\kern-.2em\hbox{\rm H}}}}
\def\IC{{\ \hbox{{\rm I}\kern-.6em\hbox{\bf C}}}}
\def\IZ{{\hbox{{\rm Z}\kern-.4em\hbox{\rm Z}}}}
\def\rref#1{(\ref{#1})}
\newcommand{\beq}{\begin{equation}}
\newcommand{\eeq}{\end{equation}}
\newcommand{\PLB}[1]{{\sl Phys.~Lett.}~{\bf B#1}}

\newcommand{\PRD}[1]{{\sl Phys.~Rev.}~{\bf D#1}}

\def \TeX{T\kern-.1667em\lower.5ex\hbox{E}\kern-.125em X}

\def \om {\omega}

\def \th {\theta}
\def \la {\lambda}
\def \La {\Lambda}
\def \Da {\Delta}

\def \a {\alpha}

\def \ga {\gamma}

\def \da {\delta}
\def \ep {\epsilon}
\def \part {\partial}

\def \um {{1\over 2}}

\def \Tr {\hbox{Tr}}

\def \sqr#1#2{{\vcenter{\hrule height.#2pt
       \hbox{\vrule width.#2pt height#1pt \kern#1pt
          \vrule width.#2pt}
       \hrule height.#2pt}}}

\def\lsim{\mathrel{\rlap{\lower4pt\hbox{\hskip1pt$\sim$}}
    \raise1pt\hbox{$<$}}}         
\def\gsim{\mathrel{\rlap{\lower4pt\hbox{\hskip1pt$\sim$}}
    \raise1pt\hbox{$>$}}}         

\def\IR{{\hbox{{\rm I}\kern-.2em\hbox{\rm R}}}}
\def\IH{{\hbox{{\rm I}\kern-.2em\hbox{\rm H}}}}
\def\IC{{\ \hbox{{\rm I}\kern-.6em\hbox{\bf C}}}}
\def\IZ{{\hbox{{\rm Z}\kern-.4em\hbox{\rm Z}}}}
\def\delbar{\,\overline{\mathop{\!\nabla\!}}\,}
\begin{document}
%
%
%
%
\def\citen#1{%
\edef\@tempa{\@ignspaftercomma,#1, \@end, }
\edef\@tempa{\expandafter\@ignendcommas\@tempa\@end}%
\if@filesw \immediate \write \@auxout {\string \citation {\@tempa}}\fi
\@tempcntb\m@ne \let\@h@ld\relax \let\@citea\@empty
\@for \@citeb:=\@tempa\do {\@cmpresscites}%
\@h@ld}
%
\def\@ignspaftercomma#1, {\ifx\@end#1\@empty\else
   #1,\expandafter\@ignspaftercomma\fi}
\def\@ignendcommas,#1,\@end{#1}
%
%
\def\@cmpresscites{%
 \expandafter\let \expandafter\@B@citeB \csname b@\@citeb \endcsname
 \ifx\@B@citeB\relax 
    \@h@ld\@citea\@tempcntb\m@ne{\bf ?}%
    \@warning {Citation `\@citeb ' on page \thepage \space undefined}%
 \else
    \@tempcnta\@tempcntb \advance\@tempcnta\@ne
    \setbox\z@\hbox\bgroup 
    \ifnum\z@<0\@B@citeB \relax
       \egroup \@tempcntb\@B@citeB \relax
       \else \egroup \@tempcntb\m@ne \fi
    \ifnum\@tempcnta=\@tempcntb 
       \ifx\@h@ld\relax 
          \edef \@h@ld{\@citea\@B@citeB}%
       \else 
          \edef\@h@ld{\hbox{--}\penalty\@highpenalty \@B@citeB}%
       \fi
    \else   
       \@h@ld \@citea \@B@citeB \let\@h@ld\relax
 \fi\fi%
 \let\@citea\@citepunct
}
%
\def\@citepunct{,\penalty\@highpenalty\hskip.13em plus.1em minus.1em}%
%
%
\def\@citex[#1]#2{\@cite{\citen{#2}}{#1}}%
%
%
\def\@cite#1#2{\leavevmode\unskip
  \ifnum\lastpenalty=\z@ \penalty\@highpenalty \fi 
  \ [{\multiply\@highpenalty 3 #1
      \if@tempswa,\penalty\@highpenalty\ #2\fi 
    }]\spacefactor\@m}
\let\nocitecount\relax  
%
\begin{titlepage}
\vspace{.5in}
\begin{flushright}
UCD-94-37\\
DFTT/49/94\\
gr-qc/9411031\\
November 1994\\
\end{flushright}
\vspace{.5in}
\begin{center}
{\Large\bf
 COMPARATIVE QUANTIZATIONS\\[1ex] OF (2+1)-DIMENSIONAL GRAVITY}\\
\vspace{.4in}
{S.~C{\sc arlip}\footnote{\it email: carlip@dirac.ucdavis.edu}\\
       {\small\it Department of Physics}\\
       {\small\it University of California}\\
       {\small\it Davis, CA 95616}\\{\small\it USA}}\\
\vspace{1ex}
{J.\ E.~N{\sc elson}\footnote{\it email: nelson@to.infn.it}\\
       {\small\it Dipartimento di Fisica Teorica }\\
       {\small\it Universit\`a degli Studi di Torino}\\
       {\small\it via Pietro Giuria 1, 10125 Torino}\\{\small\it Italy}}
\end{center}

\vspace{.5in}
\begin{center}
{\large\bf Abstract}
\end{center}
\begin{center}
\begin{minipage}{4.4in}
{\small We compare three approaches to the quantization of
(2+1)-dimensional gravity with a negative cosmological constant:
reduced phase space quantization with the York time slicing,
quantization of the algebra of holonomies, and quantization of
the space of classical solutions.  The relationships among these
quantum theories allow us to define and interpret time-dependent
operators in the ``frozen time'' holonomy formulation.
}
\end{minipage}
\end{center}

\end{titlepage}
\addtocounter{footnote}{-1}

\section{Introduction}

Over the past few years, there has been a growing interest in
(2+1)-dimensional quantum gravity as a simple model for realistic
(3+1)-dimensional quantum gravity.  As a generally covariant theory
of spacetime geometry, general relativity in 2+1 dimensions has the
same conceptual foundations as ordinary (3+1)-dimensional gravity.
But the reduction in the number of dimensions greatly simplifies
the structure of the theory, reducing the infinite number of physical
degrees of freedom of ordinary general relativity to a finite number
of global degrees of freedom.  The model thus allows us to explore
the conceptual problems of quantum gravity within the framework of
ordinary quantum mechanics, avoiding such issues as nonrenormalizability
associated with field degrees of freedom.

A number of different approaches to quantizing (2+1)-dimensional
general relativity have been developed recently.  These include reduced
phase space quantization with ADM variables \cite{HosNak,Mon,Fujiwara};
quantization of the space of classical solutions of the first-order
Chern-Simons theory \cite{Witten,observables,dirac,ordering}; and
quantization of the holonomy algebra \cite{NR0,NR1,NR2,NR3,NR4,NRZ,
Unruh,Tur}.  Each approach has its strengths and weaknesses.  ADM
quantization, for example, leads to states and operators with clear
physical interpretations, but depends on an arbitrary classical choice
of time slicing, breaking manifest covariance.  Quantization of the
space of solutions involves no such choice, but requires a detailed
understanding of the classical solutions.  Quantization of the holonomy
algebra is also manifestly covariant, and reveals important underlying
algebraic structures, but the physical interpretation of the resulting
operators is not at all clear.

The goal of this paper is to explore the relationships among these
three methods of quantization.  Such comparisons have been made in the
past \cite{dirac,Six,Ezawa0,Ezawa1,Anderson}, but the powerful holonomy
algebra approach has not generally been considered.  We shall see below
that quantization of the space of solutions (sometimes called ``covariant
canonical quantization'') provides a natural bridge between the ADM and
holonomy algebra approaches, allowing one to introduce ``time''-dependent
physical operators into the latter formalism.

The structure of the paper is as follows.  In section 2, we discuss
the first- and second-order formulations of classical general relativity,
solving the constraints and introducing the basic physical variables in
each approach.  In section 3, we describe the classical solutions for
spacetimes with the topology $\IR\!\times\!T^2$, focusing on the case
of a negative cosmological constant but also discussing the $\Lambda\!
\rightarrow\!0$ limit and briefly considering the $\Lambda\!>\!0$ case.
In section 4, we describe the three methods of quantization, and explore
their relationships.  Our results are summarized in section 5.

A preliminary report on aspects of this work has appeared in \cite{CarNel}.
Portions of our discussion of classical solutions and our comparison of
ADM and Chern-Simons quantization have been found independently by Ezawa
\cite{Ezawa1,Ezawa2}, who also discusses the $\Lambda\!>\!0$ case in more
detail.

\section{Classical Theories}
\setcounter{footnote}{0}

To understand the quantization of (2+1)-dimensional gravity, it is
first necessary to understand the classical theory.  Classical general
relativity has two very different formulations: the second-order form,
in which the metric is the only fundamental variable, and the first-order
form, in which the metric and the connection (or spin connection) are
treated independently.  As we shall see, these two formulations lead
naturally to two different approaches to quantization.

The fundamental feature of classical general relativity in 2+1 dimensions
is that the full Riemann curvature tensor depends linearly on the Ricci
tensor.  As a result, the empty space field equations $R_{\mu\nu}=
2\Lambda g_{\mu\nu}$ imply that spacetime has constant curvature, that
is, that every point has a neighborhood isometric to a neighborhood of
de Sitter, Minkowski, or anti-de Sitter space.  For a topologically
trivial spacetime, this condition eliminates all degrees of freedom.
For a spacetime with nontrivial topology, however, there remain a finite
number of degrees of freedom that describe the gluing of constant curvature
patches around noncontractible curves (see \cite{hololoop} for a more
detailed description).  It is these degrees of freedom that we shall
eventually quantize.

\subsection{ADM Formalism}

A most traditional approach to classical gravity in 2+1 dimensions
is to begin with the standard second-order form of the Einstein action
and perform an ADM-style splitting into spatial and temporal components.
This approach has been discussed in some detail by Moncrief \cite{Mon}
and Hosoya and Nakao \cite{HosNak}; in this section, we briefly
summarize their results.  We assume that spacetime has the topology
$\IR\!\times\!\Sigma$, where $\Sigma$ is a closed genus $g$ surface.
The Einstein action takes the form
\begin{eqnarray}
I_{\hbox{\scriptsize \it Ein}}
  &=& \int\!d^3x \sqrt{-{}^{\scriptscriptstyle(3)}\!g}\>
  ({}^{\scriptscriptstyle(3)}\!R - 2\Lambda) \nonumber\\
  &=& \int dt\int\nolimits_\Sigma d^2x \bigl(\pi^{ij}{\dot g}_{ij}
               - N^i{\cal H}_i -N{\cal H}\bigr),
\label{bb1}
\end{eqnarray}
where the metric has been decomposed as
\beq
ds^2 = N^2dt^2 - g_{ij}(dx^i + N^i dt)(dx^j + N^j dt)
\label{bb2}
\eeq
and the momenta conjugate to the $g_{ij}$ are $\pi^{ij} = \sqrt{g}\,(K^{ij}
- g^{ij}K)$, where $K^{ij}$ is the extrinsic curvature of the surface
$t={\rm const.}$\footnote{We use standard ADM notation: $g_{ij}$ and $R$
refer to the induced metric and scalar curvature of a time slice, while
the spacetime metric and curvature are denoted ${}^{\scriptscriptstyle(3)}
\!g_{\mu\nu}$ and ${}^{\scriptscriptstyle(3)}\!R$.}  The supermomentum and
super-Hamiltonian constraints are
\beq
{\cal H}_i = -2\nabla_j\pi^j_{\ i}\,, \qquad
{\cal H} = {1\over\sqrt{g}}\,g_{ij}g_{kl}(\pi^{ik}\pi^{jl}-\pi^{ij}\pi^{kl})
                 -\sqrt{g}(R - 2\Lambda) .
\label{bb3}
\eeq
A convenient coordinate choice is the York time slicing \cite{York}, in
which the mean (extrinsic) curvature is used as a time coordinate, $K =
\pi/\sqrt{g} = \tau$.  In reference \cite{Mon}, Moncrief shows that this
is a good global coordinate choice for classical solutions of the field
equations.  For spacetimes that do not satisfy the field equations,
on the other hand, $\tau$ need not be a good coordinate---there is no
{\em a priori} reason that a spacetime should not admit two different
slices with the same value of $K$.

The next step is to solve the constraints ${\cal H} = 0$, ${\cal H}^i=0$
to obtain a theory on a reduced phase space.  We begin with a convenient
parametrization of the spatial metric and momentum.  By the uniformization
theorem of Riemann surfaces \cite{Abikoff}, any two-metric on $\Sigma$ can
be written in the form
\beq
g_{ij} = e^{2\lambda}{\bar g}_{ij} ,
\label{bb4}
\eeq
where $\bar g_{ij}$ is a metric of constant curvature $k$ on $\Sigma$,
with $k=1$ for the sphere, $0$ for the torus, and $-1$ for any surface of
genus $g\ge2$.  Moreover, it is a standard result that up to spatial
diffeomorphisms, the space of such constant curvature metrics is finite
dimensional; that is, any $\bar g_{ij}$ can be obtained by a suitable
diffeomorphism from one of a finite-dimensional family of metrics
$\bar g_{ij}(m_\alpha)$.  The dimension of this space of ``standard''
metrics is $0$ for the sphere (that is, there is only one diffeomorphism
class of constant curvature $1$ metrics on $S^2$), $2$ for the torus, and
$6g-6$ for a surface of genus $g\ge2$.

For the torus, for instance, we can write $\bar g_{ij}=\bar g_{ij}(m)$,
where $m = m_1 + im_2$ is a complex number, the modulus, which can be taken
without loss of generality to have a positive imaginary part.\footnote{In
the mathematics literature, the modulus is usually denoted by $\tau$.
Following Moncrief \cite{Mon}, however, we have already used $\tau$ to
denote the York time coordinate.}  Concretely, the spatial metric
corresponding to a given value of $m$ is
\beq
d\sigma^2 = m_2^{-1}\left| dx + mdy \right|^2 ,
\label{bb5}
\eeq
where $x$ and $y$ each have period $1$.  Any other flat metric on the torus
is diffeomorphic to one of this form, up to a rescaling that can be absorbed
in $\lambda$ in equation \rref{bb4}.  For higher genus surfaces, we can
again write $\bar g_{ij}$ as a function of $6g-6$ moduli $m_\alpha$.  An
explicit representation of the metric similar to \rref{bb5} is much more
difficult, however, although it may be possible to handle the genus 2 case
by using the theory of hyperelliptic surfaces.

A similar decomposition is possible for the canonical momenta $\pi^{ij}$;
one obtains
\beq
\pi^{ij} = e^{-2\lambda}\sqrt{\bar g} \left(
  p^{ij}+ {1\over2}\bar g^{ij}\pi/\sqrt{\bar g}
  + \delbar^iY^j + \delbar^jY^i - \bar g^{ij}\delbar_kY^k \right),
\label{bb6}
\eeq
where $\delbar_i$ is the covariant derivative for the connection
compatible with $\bar g_{ij}$, indices are now raised and lowered with
$\bar g_{ij}$, and $p^{ij}$ is a transverse traceless tensor with respect
to $\delbar_i$,
\beq
\delbar_i\, p^{ij} = 0 .
\label{bb7}
\eeq
In the language of Riemann surface theory, $p^{ij}$ is a holomorphic
quadratic differential.  Roughly speaking, $p^{ij}$ is canonically
conjugate to $\bar g_{ij}$, $\pi$ to $\lambda$, and $Y^i$ to the spatial
diffeomorphisms.  Standard mathematical results guarantee that our
dimensions match---the dimension of the space of quadratic differentials
$p^{ij}$ is equal to the dimension of the space of diffeomorphism classes
of constant curvature metrics $\bar g_{ij}$, and, in fact, the quadratic
differentials naturally parametrize the cotangent space of the moduli space
\cite{Abikoff}.

With these parametrizations of $g_{ij}$ and $\pi^{ij}$, the momentum
constraints simply imply that $Y^i=0$, while the Hamiltonian constraint
becomes
\beq
{\cal H} = -{1\over2}\sqrt{\bar g}e^{2\lambda}(\tau^2 - 4\Lambda)
  + \sqrt{\bar g}e^{-2\lambda} p^{ij}p_{ij} + 2\sqrt{\bar g}\left[
  \bar\Delta\lambda - {1\over2}\bar R \right] = 0 .
\label{bb8}
\eeq
Moncrief has shown that this constraint uniquely determines $\lambda$
as a function of the $\bar g_{ij}$ and $p^{ij}$.  If we define a set
of coordinates $p^\alpha$ conjugate to the moduli $m_\alpha$ by
\beq
p^\alpha = \int_\Sigma e^{2\lambda}\,
  \pi^{ij}{\partial\bar g_{ij}\over\partial m_\alpha}\,d^2x ,
\label{bb9}
\eeq
the action \rref{bb1} reduces to
\beq
I_{\hbox{\scriptsize \it Ein}}
  = \int d\tau \left( p^\alpha {dm_\alpha\over d\tau} - H(m,p,\tau) \right)
\label{bb10}
\eeq
with
\beq
H = \int_\Sigma \sqrt g\,d^2x
  = \int_\Sigma e^{2\lambda(m,p,\tau)}\sqrt{\bar g}\,d^2x ,
\label{bb11}
\eeq
where $\lambda(m,p,\tau)$ is determined by \rref{bb8}.  Three-dimensional
gravity is thus reduced to a finite-dimensional mechanical system, albeit
one with a complicated and time-dependent Hamiltonian.  In particular,
the symplectic structure on the reduced phase space $(m_\alpha,p^\alpha)$
can be read off from the action \rref{bb10}:
\beq
\left\{ m_\alpha, p^\beta \right\} = \delta^\beta_\alpha .
\label{bb12}
\eeq

This system---and in particular the Hamiltonian \rref{bb11}---is generally
quite complicated, but it simplifies greatly when $\Sigma$ is a torus.
In that case, the super-Hamiltonian constraint \rref{bb8} requires that
$\lambda$ be spatially constant, and the Hamiltonian reduces to
\beq
H = \left(\tau^2-4\Lambda\right)^{-1/2}\left[m_2{}^2 p\bar p\right]^{1/2} .
\label{bb13}
\eeq
The momentum-dependent term in this expression may be recognized as the
square of the momentum $p$ with respect to the Poincar\'e (constant
negative curvature) metric
\beq
{dm d\bar m \over m_2{}^2}
\label{bb13a}
\eeq
that is the standard metric on the torus moduli space.  The Poisson
brackets \rref{bb12} are now
\beq
\left\{ m, \bar p\right\} = \left\{ \bar m, p\right\} = 2 ,\quad
\left\{ m, p\right\} = \left\{ \bar m, \bar p\right\} = 0.
\label{bb14}
\eeq

By construction, the moduli $m_\alpha$ and momenta $p^\alpha$ are
invariant under infinitesimal diffeomorphisms, and therefore under
diffeomorphisms that can be obtained by exponentiating such infinitesimal
transformations.  For a spacetime with the topology $\IR\!\times\!\Sigma$,
however, there are also ``large'' diffeomorphisms, which cannot be
obtained in this fashion.  These are generated by Dehn twists, that is,
by the operation of cutting open a handle, twisting one end by $2\pi$,
and regluing the cut edges.  The set of equivalence
classes of such large diffeomorphisms (modulo diffeomorphisms that can
be deformed to the identity) is known as the mapping class group of
$\Sigma$; for the torus, it is also known as the modular group.

For the torus, in particular, there are two independent Dehn twists,
corresponding to the two independent circumferences $\gamma_1$ and
$\gamma_2$, which we choose to have intersection number $+1$.  These
act on the fundamental group $\pi_1(T^2)$ by interchanging or mixing
the circumferences $\gamma_1$ and $\gamma_2$:
\begin{eqnarray}
&S&:\ga_1\rightarrow \ga_2^{-1},\hphantom{\ga_2}
    \qquad \ga_2\rightarrow\ga_1\nonumber\\
&T&:\ga_1\rightarrow\ga_1\cdot\ga_2 ,\qquad\ga_2\rightarrow\ga_2 ,
\label{bb16}
\end{eqnarray}
where the dot in the last line of \rref{bb16} represents composition
of curves, or multiplication of homotopy classes.  These transformations
do not leave the modulus invariant, but instead give rise to the modular
transformations
\begin{eqnarray}
&S&: m\rightarrow -{1\over m} ,\hphantom{1}\qquad
     p\rightarrow \bar m^2 p \nonumber\\
&T&: m\rightarrow m+1 ,\qquad p\rightarrow p ,
\label{bb15}
\end{eqnarray}
which may be seen to preserve the Poincar\'e metric \rref{bb13a} and
the Poisson brackets \rref{bb14}.  It may be shown that the
transformations \rref{bb16} generate the entire group of large
diffeomorphisms of $\IR\!\times\!T^2$.

Classically, observables should presumably be invariant under all spacetime
diffeomorphisms, including those in the mapping class group.  Quantum
mechanically, this condition may be relaxed, but operators and wave
functions should still transform under some unitary representation of
the mapping class group.  This restriction will be important when we
discuss quantization.

\subsection{First-Order Formalism}

Rather than starting with the metric as the fundamental variable, we
may instead write the Einstein action in first-order form, treating the
triad one-form (or coframe)  $e^a = e^a{}_\mu dx^\mu$ and the spin
connection $\omega^{ab}=\omega^{ab}{}_\mu dx^\mu$ as independent
variables.\footnote{Note that $\omega^{\hbox{\scriptsize\it here}}=
-\omega$ of \cite{Witten}.}  This leads to the first-order, connection
approach to (2+1)-dimensional gravity, inspired by Witten \cite{Witten}
(see also \cite{Achu}) and developed by Nelson, Regge and Zertuche
\cite{NR0,NR1,NR2,NR3,NR4,NRZ}.  The triad $e^a$ is related to the metric
of the previous section through
\beq
g_{\mu \nu} = {e^a}_{\mu} {e^b}_{\nu} \eta_{ab} ,
\label{bc0}
\eeq
and the action \rref{bb1} becomes
\beq
I_{\hbox{\scriptsize \it Ein}}
  = \int\- (d\omega^{ab}-{\omega^a}_d \wedge\omega^{db}
  +{\Lambda \over 3} e^a\wedge e^b)\wedge e^c\,\epsilon_{abc} ,
\qquad a,b,c=0,1,2.
\label{b2}
\eeq
For $\Lambda\ne0$, this action can be written (apart from a total
derivative) in the Chern-Simons form
\beq
I_{\hbox{\scriptsize CS}} = - {\a \over 4}
\int(d\omega^{AB}-{2\over3}\omega^A{}_E\wedge\omega^{EB})
\wedge\omega^{CD} \epsilon_{ABCD} ,\qquad A,B,C = 0,1,2,3
\label{b3}
\eeq
provided the (anti-)de Sitter spin connection $\omega^{AB}$ is identified
with the variables $e^a$, $\omega^{ab}$ in the following way:

Let $k$ denote the sign of $\Lambda$ (i.e., $k=+1$ for de Sitter space
and $k=-1$ for anti-de Sitter space), and set
$$
\Lambda =  k\alpha^{-2} .
$$
Let $\sqrt k$ mean unambiguously $+1$ for $k=1$ and $+i$ for $k=-1$.
Define the tangent space metric as $\eta_{AB}=(-1,1,1,k)$ and the
Levi-Civita density as $\epsilon_{abc3}=-\epsilon_{abc}$.  Now incorporate
the triads by setting $e^a=\a\omega^{a3}$, that is,
\beq
{\omega^A}_B=\left( \begin{array}{cc}
\omega^a{}_b& {k\over\a}e^a\\[1ex] -{1\over\a}e^b & 0 \end{array} \right) .
\label{bc1}
\eeq

The curvature two-form for the connection $\omega^{AB}$,
\beq
R^{AB}=d\omega^{AB}-\omega^{AC}\wedge\omega_C{}^B ,
\label{b4}\eeq
has components $R^{ab}+\La e^a \wedge e^b$, $R^{a3}={1\over\a}R^a$, with
\begin{eqnarray}
R^{ab} &=& d\omega^{ab} - \omega^{ac}\wedge
  \omega_{c}{}^{b} \nonumber\\
R^{a} &=& de^a -\omega^{ab}\wedge e_b .
\label{bc2}
\end{eqnarray}
$R^{ab}$ and $R^a$ may be recognized the ordinary (2+1)-dimensional
curvature and torsion forms.  The field equations derived from the
action \rref{b3} are simply
\beq
R^{AB}=0 ,
\label{b44}
\eeq
implying, as in the second-order
formalism of the previous section, that the torsion vanishes everywhere
and that the curvature $R^{ab}$ is constant.

In a (2+1)-dimensional splitting of spacetime, the action \rref{b3}
decomposes as
\beq
I_{\hbox{\scriptsize CS}} =
{\a\over 4} \int\!dt\int\!d^2x\,\epsilon^{ij}\epsilon_{ABCD}\,
 (\omega^{CD}{}_j\,{\dot{\omega}}^{AB}{}_i-\omega^{AB}{}_0 R^{CD}{}_{ij})
\label{b5}
\eeq
(with $\epsilon^{0ij} = -\epsilon^{ij}$), from which the constraints are
\beq
R^{AB}{}_{ij}=0 .
\label{b8}
\eeq
These are equivalent to the conditions ${\cal H}=0$, ${\cal H}^i=0$ of
equation \rref{bb3} of the last section.  The Poisson brackets can be
read off from \rref{b5}:
\beq
\{\omega^{AB}{}_{i}(x),\omega^{CD}{}_{j}(y)\}
 ={k\over2\a}\epsilon_{ij} \epsilon^{ABCD}\da^{2}(x-y) ,
\label{b6}
\eeq
or equivalently
\begin{eqnarray}
\{ e^a{}_i({\bf x}),\omega^{bc}{}_j({\bf y}) \}
 &=& -\um \epsilon_{ij}\epsilon^{abc}\da^2({\bf x-y})\nonumber\\
\{ e^a{}_i(x),e^b{}_j(y) \}&=&\{ \omega^{ab}{}_i(x),\omega^{cd}{}_j(y) \}=0
\label{bc2a}
\end{eqnarray}
where $x,y\in\Sigma$ are generic points on the $t=\hbox{const.}$
surface $\Sigma$ and $\epsilon_{12}=1$.

The constraints \rref{b8} imply that the (anti-)de Sitter connection
$\omega^{AB}{}_i$ is flat.  It can therefore be written locally in
terms of an $\hbox{SO}(3,1)$- or $\hbox{SO}(2,2)$-valued zero-form
$\psi^{AB}$ as
\beq
d\psi^{AB}=\om^{AC}\, \psi_C{}^B .
\label{bc3}
\eeq
Of course, $\omega^{AB}$ may have nontrivial holonomies around closed
loops in $\Sigma$, so $\psi^{AB}$ will not necessarily be single-valued.
It is actually more convenient to use the spinor groups, where the spinor
group of $\hbox{SO}(2,2)$ is $\hbox{SL}(2,\IR)\otimes \hbox{SL}(2,\IR)$
and that of $\hbox{SO}(3,1)$ is $\hbox{SL}(2,\IC)$.  Define the one-form
\beq
\Da(x) = \Da_{i}(x)dx^{i} = {1\over 4}\omega^{AB}(x)\ga_{AB}
\label{bc4}
\eeq
(for $\ga$-matrix conventions and identities see the Appendix), for which
\rref{b8} implies that $d\Da-\Da\wedge\Da=0$.  This form of the constraints
can be integrated using multivalued $\hbox{SL}(2,\IR)$ or $\hbox{SL}(2,\IC)$
matrices $S$, which satisfy
\beq
dS(x)=\Da(x)S(x) .
\label{bc5}
\eeq
These are related to the matrices $\psi$ of \rref{bc3} by
\beq
\psi^{AB}\ga_{B}=S^{-1}\ga^{A}S .
\eeq

{}From \rref{b6}, these spinor representations satisfy the Poisson brackets
\begin{eqnarray}
\{\Da_{i}^{\pm}(x), \Da_{j}^{\pm}(y)\}
 &=& \pm {i\over2\a\sqrt{k}}\ep_{ij}\sigma^{m}\otimes\sigma^{m}\da^{2}(x-y)
 \nonumber\\
\{\Da_{i}^{+}(x), \Da_{j}^{-}(y)\} &=& 0  ,
\end{eqnarray}
where the $\sigma^{m}$ are Pauli matrices and the $\pm$ refer to the
decomposition of the $4\times 4$ representations of $\Da(x)$, $S(x)$ into
$2\times 2$ irreducible parts by using the projectors $p_{\pm}$
(see Appendix),
\begin{eqnarray}
\Da(x) &=& \Da^{+}(x)\otimes p_{+}+\Da^{-}(x)\otimes p_{-} \nonumber\\
S(x) &=& S^{+}\otimes p_{+}+S^{-}(x)\otimes p_{-} ,
\end{eqnarray}
so
\beq
dS^{\pm}(x) = \Da^{\pm}(x)S^{\pm}(x) .
\eeq

In this approach to (2+1)-dimensional gravity, the existence of
nontrivial classical solutions arises from the fact that the $S^\pm$
may be multivalued---that is, the connections $\omega^{AB}$ and $\Delta$
may have nontrivial holonomies.  In particular, let $\gamma\!:\,[0,1]\!
\rightarrow\!\Sigma$ be a noncontractible closed curve based at
$\gamma(0) = x_0$, and take $S^\pm(\gamma(0)) = 1$ as an initial
condition for the differential equation \rref{bc5}.  Then \rref{bc5}
can be integrated to obtain a nontrivial value for $S^\pm(\gamma(1))
= S^\pm[\gamma]$, the $\hbox{SL}(2,\IR)$ or $\hbox{SL}(2,\IC)$ holonomy
of $\Delta$.  In general, this holonomy will depend on the curve
$\gamma$, but for our system, the flatness of the connection $\Delta$
implies that $S^\pm[\gamma]$ depends only on the homotopy class of
$\gamma$.

The Poisson brackets \rref{b6} now induce corresponding brackets between
$S^{\pm}[\sigma]$ and $S^{\pm}[\gamma]$, where $\sigma,\gamma\!\in\!
\pi_{1}(\Sigma,x_0)$. The matrices $S^{\pm}[\gamma]$ therefore furnish
a representation of $\pi_{1}(\Sigma,x_0)$ in $\hbox{SL}(2,\IR)$ or
$\hbox{SL}(2,\IC)$.  Under a gauge transformation or a change of base
point, the $S^\pm$ transform by conjugation, so their traces provide
an (overcomplete) set of gauge-invariant variables.

The classical Poisson brackets for these variables were calculated by
hand for the genus $1$ and genus $2$ cases, and then generalized and
quantized in \cite{NR1}.  For the genus $1$ case that is the
focus of this paper, the Poisson algebra is
\beq
\{R_1^{\pm},R_2^{\pm}\}=\mp{i\over {4\a\sqrt k}}(R_{12}^{\pm}-
 R_1^{\pm}R_2^{\pm}) \quad \hbox{\it and cyclical permutations},
\label{b7}
\eeq
where $R^{\pm}= \um \hbox{Tr}S^{\pm}$.  Here the subscripts $1$ and $2$
refer to the two independent intersecting circumferences $\gamma_1$, $\gamma_2$
on $\Sigma$ with intersection number $+1$,\footnote{Paths with intersection
number 0, $\pm$ 1 are sufficient to characterize the holonomy algebra for
genus $1$.  For $g>1$, one must in general consider paths with two or more
intersections, for which the brackets \rref{b7} are more complicated; see
\cite{NR3,NR4}.} while the third holonomy, $R^\pm_{12}$, corresponds to
the path $\gamma_1\cdot\gamma_2$, which has intersection number $-1$ with
$\gamma_1$ and $+1$ with $\gamma_2$.

Classically, the six holonomies $R^\pm_{1,2,12}$ provide an overcomplete
description of the spacetime geometry of $\IR\!\times\!T^2$, which, as
we saw in the last section, is completely characterized by four real or
two complex parameters $m$ and $p$.  To understand this overcompleteness,
consider the cubic polynomials
\beq
F^{\pm}=1-(R_1^{\pm})^2-(R_2^{\pm})^2-(R_{12}^{\pm})^2 +
 2 R_1^{\pm}R_2^{\pm}R_{12}^{\pm} .
\label{b9}
\eeq
These polynomials have vanishing Poisson brackets with all of the
traces $R_a^{\pm}$, and are cyclically symmetric in the $R_a^{\pm}$.
Moreover, the classical algebra \rref{b7} and the central elements
\rref{b9} are invariant under the modular transformations \rref{bb16}.

The overcompleteness of our description arises because the polynomials
$F^\pm$ vanish classically by the $\hbox{SL}(2,\IR)$ or $\hbox{SL}(2,\IC)$
Mandelstam identities. Indeed, $F^\pm$ may be expressed as
\beq
F^{\pm}= \um\, \hbox{Tr}\left(I - S^{\pm}[\ga_1]S^{\pm}[\ga_2]S^{\pm}
[\ga_1^{-1}]S^{\pm}[\ga_2^{-1}]\right) ,
\label{b10}
\eeq
where $S^{\pm}[\ga_i^{-1}]=(S^{\pm}[\ga_i])^{-1}$; this expression reduces
to \rref{b9} by virtue of the identities
$$
A + A^{-1} = I\,\hbox{Tr}A
$$
for $2 \times 2$ unimodular matrices $A$.  The classical vanishing
of $F^{\pm}$ can thus be thought of as the application of the group
identity
$$
\ga_1^{\vphantom{-1}}\cdot\ga_2^{\vphantom{-1}}\cdot\ga_1^{-1}\cdot\ga_2^{-1}
  = I
$$
to the representations $S^{\pm}$.

In this first-order approach, the constraints have now been solved exactly.
There is no Hamiltonian, however, and no time development. One can think
of this formalism as initial data for some (unspecified) choice of time,
or alternatively as giving a time-independent description of the entire
spacetime geometry.

\section{Classical Solutions}
\setcounter{footnote}{0}

Before turning to quantization, it is useful to explore the structure
of the classical solutions of (2+1)-dimensional gravity in more detail.
We shall concentrate on spacetimes with the topology $\IR\!\times\!T^2$,
for which the space of classical solutions is completely understood.  It
would, of course, be valuable to extend these results to more complicated
topologies.  Some progress towards this goal has been made in the holonomy
formulation \cite{NR0}, but a full geometric understanding is still lacking.
We shall also specialize to the case $\Lambda\!<\!0$, briefly discussing the
corresponding picture for $\Lambda\!\ge\!0$ at the end of this section.  Many
of the results presented in this section have been discovered independently
by Ezawa \cite{Ezawa1,Ezawa2}, and related solutions were found by Fujiwara
and Soda \cite{Fujiwara}.

An obvious place to begin is with the ADM formalism of section 2.1: the
spatial metric on a constant $\tau$ slice is given by equation \rref{bb5},
and the results of Moncrief \cite{Mon} permit a straightforward computation
of the lapse and shift functions in the full ADM metric \rref{bb2}.  Rather
than using the York time slicing from the start, it is actually somewhat
easier to begin in a slightly different gauge, ``time gauge,'' in which
$N=1$ and $N^i=0$.  Equivalently, in the first order form, we choose
$e^0_i=0$ and $e^0_t=1$.  We shall see below that for the topology
$\IR\!\times\!T^2$, this choice is equivalent to the York gauge, although
this is no longer the case for spaces of genus $g\!>\!1$.

With this gauge choice, it is easy to check that the first-order field
equations \rref{b44} are solved by
\begin{eqnarray}
e^0 &=&dt \nonumber\\
e^1&=& {\a \over 2} \left[(r_1^+ - r_1^-)dx +(r_2^+ - r_2^-)dy\right]
 \sin{t \over {\a}} \\
e^2&=& {\a \over 2} \left[(r_1^+ + r_1^-)dx +(r_2^+ + r_2^-)dy\right]
 \cos{t \over {\a}} \nonumber
\label{c1}
\end{eqnarray}
\begin{eqnarray}
\om^{12}&=&0  \nonumber\\
\om^{01}&=& -{1 \over 2} \left[(r_1^+ - r_1^-)dx +(r_2^+ - r_2^-)dy\right]
 \cos{t \over {\a}} \\
\om^{02}&=& {1 \over 2} \left[(r_1^+ + r_1^-)dx +(r_2^+ + r_2^-)dy\right]
 \sin{t \over {\a}} , \nonumber
\label{c2}
\end{eqnarray}
where $r_1^\pm$ and $r_2^\pm$ are four arbitrary parameters and the
coordinates $x$ and $y$ have period one.  The triad \rref{c1} determines
a spacetime metric $g_{\mu\nu} = e^a{}_\mu e_{a\nu}$, which can be used
to compute the moduli and momenta of section 2.1.  In particular, it is
not hard to show that the spatial metric $g_{ij}$ on a slice of constant $t$
describes a torus with modulus
\beq
m= \left(r_1^-e^{it/\a} + r_1^+e^{-{it/\a}}\right)
 \left(r_2^-e^{it/\a} + r_2^+e^{-{it/\a}}\right)^{\lower2pt%
 \hbox{$\scriptstyle -1$}} .
\label{c3}
\eeq
The conjugate momentum $p$ of equation \rref{bb9} can be similarly computed
from the $g_{\mu\nu}$ in terms of the extrinsic curvature of a slice of
constant $t$; it takes the form
\beq
p= -{i\a\over 2\sin{2t\over \a}}\left(r_2^+e^{it/\a}
 + r_2^-e^{-{it/\a}}\right)^{\lower2pt%
 \hbox{$\scriptstyle 2$}} .
\label{c4}
\eeq
Finally, the York time is
\beq
\tau = -{d\over dt}\ln\sqrt{g} = -{2\over\alpha}\cot{2t\over\alpha} ,
\label{c5}
\eeq
which ranges from $-\infty$ to $\infty$ as $t$ varies from $0$ to
$\pi\alpha/2$.  Clearly, $\tau$ is a monotonic function of $t$ in this
range, so a slicing by surfaces of constant $t$ is equivalent to the
York slicing by surfaces of constant $K$, as claimed.

To check the generality of the solution \rref{c1}--\rref{c2}, observe
first that the four parameters $r_a^\pm$ can be chosen arbitrarily,
which in turn implies that the modulus $m$ and momentum $p$ of equations
\rref{c3} and \rref{c4} can take arbitrary values at an initial surface
$t=t_0$.  This means that we can specify arbitrary initial data $(m(t_0),
p(t_0))$ in the ADM formalism.  Results of Moncrief \cite{Mon} and Mess
\cite{Mess} then guarantee that such data determine a unique maximal
spacetime---technically, a maximal domain of dependence of the initial
surface---and that any such spacetime can be obtained from suitable
initial data.\footnote{This is no longer true in the case $\Lambda > 0$
\cite{Mess}; the resulting ambiguity is discussed briefly in
\cite{Witten:complex} and \cite{Ezawa2a}.  Moreover, as Louko and Marolf
have observed \cite{Louko}, if one starts with a solution that is a domain
of dependence, it may be possible to find further extensions to regions
containing closed timelike curves.}

We can obtain additional information about this solution by calculating
the $\hbox{SL}(2,\IR)$ holonomies of equation \rref{bc5}, using the
decomposition of the spinor group of $\hbox{SO}(2,2)$ described in
Section 2.2.  The computation is again straightforward, and gives traces
\begin{eqnarray}
R_1^\pm &=& {1\over2}\Tr S^\pm[\gamma_1] = \cosh{r_1^\pm\over2} ,
  \nonumber\\
R_2^\pm &=& {1\over2}\Tr S^\pm[\gamma_2] = \cosh{r_2^\pm\over2} ,\\
R_{12}^\pm &=& {1\over2}\Tr S^\pm[\gamma_1\cdot\gamma_2]
  = \cosh{(r_1^\pm+r_2^\pm)\over2} . \nonumber
\label{c6}
\end{eqnarray}
Conversely, the metric $g_{\mu\nu}$ can be obtained directly from
the holonomies by a quotient space construction.  Three-dimensional
anti-de Sitter space is naturally isometric to the group manifold of
$\hbox{SL}(2,\IR)$.  Indeed, anti-de Sitter space can be represented as
the submanifold of flat $\IR^{2,2}$ (with coordinates $(X_1,X_2,T_1,T_2)$
and metric $dS^2 = dX_1^2 + dX_2^2 - dT_1^2 - dT_2^2$) on which
\beq
\hbox{det}|{\bf X}| = 1 ,
  \qquad {\bf X} = {1\over\alpha}\left( \begin{array}{cc}
  X_1+T_1 & X_2+T^2\\ -X_2+T_2 & X_1-T_1 \end{array} \right) ,
\label{c7}
\eeq
i.e., ${\bf X}\in\hbox{SL}(2,\IR)$.  If one allows the $S_a^+$ to act on
$\bf X$ by left multiplication and the $S_a^-$ to act by right multiplication,
it may be shown that the triad \rref{c1} represents the geometry of the
quotient space $\langle S^+_1,S^+_2 \rangle\backslash\hbox{AdS}/\langle
S^-_1,S^-_2 \rangle$.

Now, the holonomies \rref{c6} are {\em not} the most general possible: an
$\hbox{SL}(2,\IR)$ matrix can have an arbitrary trace, while our solution
requires the holonomies to be hyperbolic.\footnote{An $\hbox{SL}(2,\IR)$
matrix $R$ is called hyperbolic if $|\Tr R|>2$, parabolic if $|\Tr R|=2$,
and elliptic if $|\Tr R|<2$.}  The solution \rref{c1}--\rref{c2} thus
represents only one sector in the space of holonomies, the
``hyperbolic-hyperbolic'' sector, out of nine possibilities \cite{Ezawa2}.
On the other hand, we argued above that \rref{c1}--\rref{c2} gave the
most general solution to the problem of evolution of initial data on a
spacelike surface with the topology $T^2$.  These two statements are not,
in fact, inconsistent: for solutions with elliptic or parabolic holonomies,
spacetime still has the topology $\IR\!\times\!T^2$, but the toroidal
slices are not spacelike \cite{Ezawa2}.  In particular, the choice of
time gauge is only possible when all holonomies are hyperbolic.  A similar
phenomenon has been investigated in detail by Louko and Marolf \cite{Louko}
in the case $\Lambda\!=\!0$.

\subsection{Classical Time Evolution}

By construction, we know that the traces \rref{c6} must satisfy the
nonlinear classical Poisson bracket algebra \rref{b7}.  One may easily
verify from \rref{c6} that the central elements \rref{b9} are identically
zero. The algebra \rref{b7} implies that the holonomy parameters
$r^{\pm}$ satisfy
\beq
\{r_1^\pm,r_2^\pm\}=\mp {1\over\a}, \qquad \{r^+_a,r^-_b\}=0 .
\label{a1}
\eeq
It is easily checked that the brackets \rref{a1} induce the correct
brackets \rref{bb14} for the modulus $m$ and momentum $p$ defined by
\rref{c3}--\rref{c4}, confirming the consistency of the classical first-
and second-order descriptions.

{}From \rref{c3}, it may be shown that the time-dependent moduli $m_1,m_2$
lie on a semicircle of radius $R$,
\beq
\left(m_1 -c\right)^2 + m_2^2 = R^2 ,
\eeq
where
\beq
c = {{r_1^+ r_2^+ - r_1^- r_2^-}\over{(r_2^+)^2 - (r_2^-)^2}},\quad
  R^2 =\left({{r_1^+ r_2^- - r_1^- r_2^+}\over
 {(r_2^+)^2 - (r_2^-)^2}}\right)^2 .
\eeq
This agrees with the results of Fujiwara and Soda \cite{Fujiwara}.  The
imaginary part $m_2$ of the modulus ranges from zero to $R$
and back to zero, while $m_1$ ranges from $c+R$ to $c$ to $c-R$.  The
momenta \rref{c4} are similarly constrained
through
\beq
p(\bar m - c)+ \bar p(m - c) =0 .
\eeq
In this range $p^1$ is constant while $p^2$ varies from $-\infty$ to
zero to $+\infty$.  This behavior clearly illustrates the
nontrivial dynamics of the system, even though the full (2+1)-dimensional
curvature tensor is everywhere constant.  This is {\em not} a ``gauge''
effect, but rather reflects the nontrivial, time-dependent identifications
needed to construct a torus from patches of anti-de Sitter space.

The Hamiltonian $H$ of equation \rref{bb13} generates this classical
$\tau$ development of the moduli and their momenta through Hamilton's
equations,
\beq
{dp\over d\tau}=\{p,H\},\qquad {dm\over d\tau}=\{m,H\} .
\eeq
$H$ may be calculated in terms of the holonomy parameters $r_a^{\pm}$ as
\begin{eqnarray}
H=g^{1/2}&=&{\a^2\over 4}\sin{2t\over \a}\,(r_1^-r_2^+ - r_1^+r_2^-)
  \nonumber\\
&=&{\a \over {2\sqrt{\tau^2-4\La}}}(r_1^-r_2^+ - r_1^+r_2^-) ,
\label{a2}
\end{eqnarray}
where we assume that $r_1^-r_2^+ - r_1^+r_2^-\!>\!0$ so that $g^{1/2}\!>\!0$
in the range $t\!\in\!(0,\pi\a/2)$.  This guarantees that the imaginary part
$m_2$ of the modulus is always positive, as it is in the standard description
of torus geometry.  The area of the torus expands from zero to a maximum
at $t\!=\!\pi\a/4$, $\tau\!=\!0$ and then recollapses back to zero at
$t\!=\!\pi\a/2$, $\tau\!=\!\infty$.  Alternatively, the Hamiltonian
\beq
H^\prime={d\tau\over dt}H={4\over {\a^2}}\csc^2 {2t\over \a}H
\label{a3}
\eeq
generates evolution in coordinate time $t$ by
\beq
{dp\over dt}=\{p, H^\prime\},\qquad {dm\over dt}=\{m, H^\prime\} .
\eeq

The standard action of the modular group on the traces \rref{c6}
suggests that the holonomy parameters transform as
\begin{eqnarray}
&S&:r_1^{\pm}\rightarrow r_2^{\pm},\hphantom{+ r_2^\pm r}\!\qquad
    r_2^{\pm}\rightarrow - r_1^{\pm}\nonumber\\
&T&:r_1^{\pm}\rightarrow r_1^{\pm} + r_2^{\pm},\qquad
    r_2^{\pm}\rightarrow r_2^{\pm}
\label{xa00}
\end{eqnarray}
It may be checked that these transformations do indeed leave the brackets
\rref{a1} and the Hamiltonians \rref{a2} and \rref{a3} invariant, and that
they induce the correct modular transformations \rref{bb15} on the moduli
and momenta.

The relationships among the moduli and the holonomy parameters $r^\pm$
allows us to write the reduced phase space action of (2+1)-dimensional
gravity in several equivalent forms:
\begin{eqnarray}
I_{\hbox{\scriptsize \it Ein}} = \int\!dt \int\!d^2x\,\pi^{ij} {\dot g}_{ij}
&=&\int\!dt\int\!d^2x\,2\ep^{ij}\ep_{abc}\,e^c_j\,{\dot{\om}}^{ab}_i\nonumber\\
&=&\int\um (\bar p dm + p d\bar m) + H d\tau -d(p^1m_1 +p^2m_2) \\
&=&\a \int (r_1^-dr_2^- - r_1^+ dr_2^+) ,\nonumber
\label{xa0}
\end{eqnarray}
showing that the
holonomy parameters $r_{1,2}^{\pm}$ are related to the modulus $m$
and momentum $p$ through a (time-dependent) canonical transformation.

\subsection{ $\La \to 0^-$}

Much of the classical behavior discussed above was studied previously
in \cite{HosNak,observables} for the case of vanishing cosmological
constant.  The $\Lambda=0$ theory is easy to describe in ADM variables,
but the holonomies analogous to $R^\pm_a$ are considerably more complicated,
since the relevant gauge group is the non-semisimple Lie group $\hbox{ISO}
(2,1)$, the (2+1)-dimensional Poincar\'e group.  It is therefore useful to
describe the relationship between the $\Lambda\!<\!0$ and $\Lambda\!=\!0$
theories.  (The corresponding limit for $\Lambda\!\to\!0^+$  has also
been studied by Ezawa \cite{Ezawa1}.)

The $\Lambda\!\rightarrow\!0$ limit is most easily seen by rescaling the
holonomy parameters as follows.  Define
\begin{eqnarray}
w_a&=&\a(r_a^+ + r_a^-)/2\nonumber\\
u_a&=&(r_a^+ - r_a^-)/2, \quad a=1,2
\label{p1}
\end{eqnarray}
where the (time-independent) $w_a$ and $u_a$ remain finite as $\Lambda\to0$,
$\alpha \to \infty$.  In this limit, the York time \rref{c5} becomes
$$\tau =-{1 \over t} ,$$
and the solution \rref{c1}--\rref{c2} reduces to
\begin{eqnarray}
e^0 &=&dt \nonumber\\
e^1&=& t \left[u_1dx +u_2dy\right]\\
e^2&=&  \left[w_1dx +w_2dy\right] \nonumber
\label{cb1}
\end{eqnarray}
\begin{eqnarray}
\om^{12}&=&0  \nonumber\\
\om^{01}&=& - \left[u_1dx +u_2dy\right] \\
\om^{02}&=& 0 .\nonumber
\label{cb2}
\end{eqnarray}
The moduli and momenta of \rref{c3}--\rref{c4} are now
\begin{eqnarray}
m&=&(w_1-itu_1)(w_2-itu_2)^{-1}\nonumber\\
p&=& -{i \over t}(w_2-itu_2)^2 ,
\end{eqnarray}
and the Hamiltonian \rref{a2} is
\beq
H=g^{1/2}=t (w_1u_2 - u_1w_2) .
\eeq
The recontraction of the spatial slices now disappears; the tori instead
expand linearly in the range $t\in(0,\infty)$.  The new variables $u$ and
$w$ may be easily shown to satisfy the classical Poisson brackets
\beq
\{u_1,w_2\}=\{w_1,u_2\}=- \um ,
\label{xa1}
\eeq
derivable from the action
\beq
I = -2 \int (u_1dw_2 + w_1du_2) .
\eeq
They are related to the parameters of reference \cite{observables} by
\beq
u_1= -\la,\ u_2= -\mu,\ w_1= a,\ w_1=b .
\eeq

For the traces \rref{c6} the limit is more complicated.  However,
the Poincar\'e variables and their algebra described in reference
\cite{NR2} may be retrieved by using the holonomy parameters
\beq
r_a^{\pm}= u_a \pm {w_a \over \a}
\eeq
to expand the traces \rref{c6} to first order in $1/\alpha$,
\begin{eqnarray}
R_a^{\pm} &=& \cosh{r_a^{\pm} \over 2}\nonumber\\
&=& \cosh{u_a \over 2}  \pm { w_a \over 2\a} \sinh{u_a \over 2}  \\
&=&q_a \pm {\nu_a \over \a} \qquad (a=1,2) .\nonumber
\label{p2}
\end{eqnarray}
The Poisson brackets \rref{xa1} then imply that to order $1/\alpha$,
\begin{eqnarray}
\{q_1,q_2\}&=&-{1 \over {16\a^2}}(\nu_{12}-\nu_{1{2^{-1}}})\nonumber\\
\{q_1,\nu_2\}&=&-{1 \over 16}(q_{12}-q_{1{2^{-1}}})
+{1 \over {8\a^2}}\nu_1\nu_2\\
\{\nu_1,\nu_2\}&=&-{1 \over 16}(\nu_{12}-\nu_{1{2^{-1}}}) .\nonumber
\label{p3}
\end{eqnarray}
Here
\begin{eqnarray}
q_{12}=\cosh{{u_1+u_2} \over 2}&,&
\quad q_{1{2^{-1}}}=\cosh{{u_1-u_2}\over 2}\nonumber\\
\nu_{12}=({{w_1+w_2} \over 2})\sinh{{u_1+u_2} \over 2}&,&
\quad\nu_{1{2^{-1}}}=({{w_1-w_2}\over 2})\sinh{{u_1-u_2}\over 2} ,
\label{p4}
\end{eqnarray}
which satisfy the identities
\beq
\nu_{1{2^{-1}}} +\nu_{12}=2(\nu_1 q_2 + \nu_2 q_1), \quad
q_{1{2^{-1}}} + q_{12}=2q_1 q_2 .
\label{p5}
\eeq
With these identifications, the algebra \rref{p3} reproduces that of
\cite{NR2} in the limit $\Lambda\to 0$, $\alpha\to\infty$.\footnote{Note
that the $\nu$ of \cite{NR2} is $8\nu^{here}$.}

\subsection{$\La > 0$}

The case of positive cosmological constant has been studied in detail by
Ezawa \cite{Ezawa1}. For completeness, we point out that the classical
solutions for $\Lambda\!>\!0$ can easily be derived from our solutions
for $\Lambda\!<\!0$ by substituting hyperbolic sines and cosines for
sines and cosines.  One finds that the range of
$$
\tau = -{2\over\alpha}\coth{2t\over\alpha}
$$
is now $-\infty$ to $-{2\over\a}$ for $t\in(0,\infty)$, and that the
area of the torus expands exponentially from zero to $\infty$.
The $R_a^{\pm}$ are now traces of $\hbox{SL}(2,\IC)$ holonomies, expressed
as the complex conjugates
\beq
R_a^{\pm}(r_a^+,r_a^-)= \cos{{(r_a^+ + r_a^-) \pm i(r_a^+ - r_a^-)}\over 4} ,
\label{p40}
\eeq
which can be written in terms of the parameters $u,w$ of equation \rref{p1}
of the previous section as
\begin{eqnarray}
R_a^{\pm}&=& \cos({ w_a \over 2\a} \pm i{u_a \over 2})\nonumber\\
&=&\cos({ w_a \over 2\a}) \cosh({ u_a \over 2}) \mp i\sin({ w_a \over 2\a})
\sinh({u_a \over 2})
\label{p4a}
\end{eqnarray}
To first order in $1/\alpha$, we have
\beq
R_a^{\pm} = q_a \mp i {\nu_a \over \a} \qquad (a=1,2)
\eeq
(compare with \rref{p2} for $\Lambda\!<\!0$), so the limit $\Lambda\!\to\!0$
is again easy to understand.  In particular, the $q_a$ and $\nu_a$ again
satisfy the algebra \rref{p3} in the limit $\Lambda\!\to\!0^+$.

Note that from \rref{p40},
\beq
R_a^{\pm}(r_a^+,r_a^-)
  = R_a^{\pm}(r_a^+ +4\pi n^{\phantom{+}}_a,r_a^- + 4\pi n^{\phantom{+}}_a)
\eeq
for any integers $n_a$.  Hence the parameters $r_a^{\pm}$---and therefore
the moduli---are not uniquely determined by the $\hbox{SL}(2,\IC)$ traces,
in contrast to the $\Lambda\!<\!0$ case.  Such a change corresponds to
adding the total derivative
$$4 \pi n_1 d(r_2^- -r_2^+)$$
to the action \rref{xa0}.  This ambiguity was first noted by Mess
\cite{Mess}, and was discussed by Witten in reference \cite{Witten:complex}.
It suggests that in addition to the traces $R_a^\pm$, a new discrete quantum
number related to the direct quantization of the parameters $r_a^\pm$
may be necessary to describe (2+1)-dimensional gravity with positive
cosmological constant.

\section{Quantum Theories}
\setcounter{footnote}{0}

We now turn to the quantization of the system described above.  As we shall
see, the different classical descriptions naturally lead to very different
approaches to the quantum theory, whose relationship can give us further
information about the structure of (2+1)-dimensional quantum gravity.

\subsection{ADM Quantization}

Let us begin with the second-order formalism of section 2.1.  We saw above
that the reduced phase space action \rref{bb10}---the action written in
terms of the physical variables $m_\alpha$ and $p^\alpha$---is equivalent
to that of a finite-dimensional mechanical system with a complicated
Hamiltonian.  We know, at least in principle, how to quantize such a
system: we simply replace the Poisson brackets \rref{bb12} with
commutators,
\beq
\left[ \hat m_\alpha, \hat p^\beta \right] = i\hbar\delta^\beta_\alpha ,
\label{da1}
\eeq
represent the momenta as derivatives,
\beq
p^\alpha = {\hbar\over i}{\partial\ \over\partial m_\alpha} ,
\label{da2}
\eeq
and impose the Schr\"odinger equation,
\beq
i\hbar{\partial\psi(m,\tau)\over\partial\tau} = \hat H\psi(m,\tau) ,
\label{da3}
\eeq
where the Hamiltonian $\hat H$ is obtained from \rref{bb11} by some
suitable operator ordering.

One fundamental problem, of course, is hidden in this last step: it is
not at all obvious how one should define $\hat H$ as a self-adjoint
operator on an appropriate Hilbert space.  The ambiguity is already
evident for the genus 1 Hamiltonian \rref{bb13}: $\hat m_2$ and $\hat p$
do not commute, so the operator ordering is not unique.  The simplest
choice of ordering is that of equation \rref{bb13}, for which the
Hamiltonian becomes
\beq
\hat H = {\hbar\over\sqrt{\tau^2-4\Lambda}}\Delta_0^{1/2} ,
\label{da4}
\eeq
where $\Delta_0$ is the ordinary scalar Laplacian for the constant negative
curvature moduli space characterized by the metric \rref{bb13a}.  This
Laplacian is invariant under the modular transformations \rref{bb15}; its
invariant eigenfunctions, the weight zero Maass forms, are discussed
in considerable detail in the mathematical literature \cite{Maass}.

While this choice of ordering is not unique, the number of possible
alternatives is smaller than one might fear.  The key restriction is
diffeomorphism invariance: the eigenfunctions of the Hamiltonian should
transform under a one-dimensional unitary representation of the mapping
class group.  The representation theory of the modular group \rref{bb15}
has been studied extensively \cite{Maass2}; one finds that the possible
inequivalent Hamiltonians are all of the form \rref{da4}, but with
$\Delta_0$ replaced by\footnote{It is argued in \cite{dirac} that the
natural choice of ordering in first-order quantization corresponds to
$n=1/2$.}
\beq
\Delta_n = -m_2^{\ 2}\left( {\partial^2\ \over\partial m_1^{\ 2}} +
  {\partial^2\ \over\partial m_2^{\ 2}}\right)
  + 2in\, m_2{\partial\ \over\partial m_1} + n(n+1) , \quad 2n\in\IZ ,
\label{da5}
\eeq
the Maass Laplacian acting on automorphic forms of weight $n$.  (See
\cite{ordering} for details of the required operator orderings.)  Note
that when written in terms of the momentum $p$, the operators $\Delta_n$
differ from each other by terms of order $\hbar$, as expected for operator
ordering ambiguities.  Nevertheless, the various choices of ordering can
have drastic effects on the physics: the spectra of the various Maass
Laplacians are very different.

This ambiguity can be viewed as a consequence of the structure of the
classical phase space.  The torus moduli space is not a manifold, but
rather has orbifold singularities, and quantization on an orbifold
is generally not unique.  Since the space of solutions of the Einstein
equations in 3+1 dimensions has a similar orbifold structure \cite{Isen},
we might expect a similar ambiguity in realistic (3+1)-dimensional
quantum gravity.

There is another, potentially more serious, ambiguity in this approach to
quantization, coming from the classical treatment of the time slicing.
The choice of $K$ as a time variable is rather arbitrary---it greatly
simplifies the constraints \rref{bb3}, but is otherwise no better than
any other classical gauge-fixing technique---and it is not at all clear
that a different choice would lead to the same quantum theory.  The danger
of making a ``wrong'' choice is illustrated by the classical solution
\rref{c1}--\rref{c2}: another standard gauge choice is $\sqrt{g}=t$, but
it is evident that when $\Lambda<0$, $\sqrt{g}$ is not even a single-valued
function of $\tau$.

A possible resolution of this problem is to treat the holonomy approach,
in which no choice of time slicing is needed, as fundamental.  If we can
establish a relationship between the $(\hat m,\hat p)$ and suitable
operators in the first-order formulation, we can convert the problem of
time slicing into one of defining the appropriate physical operators.
Different choices of slicing would then merely require different operators
to represent moduli, and not different quantum theories.

\subsection{Quantizing Traces of Holonomies}

We next consider an alternative approach to quantization, starting from
the first-order formulation of the classical theory.  Without assuming
{\em ab initio} any classical relationship between moduli and holonomies,
the algebra of the traces $R^\pm$ can be quantized directly for any
value of the cosmological constant $\La$ and any genus $g$ of $\Sigma$.
For arbitrary genus, one obtains an abstract quantum algebra, the subject
of intense study \cite{NR0,NR1}.  In principle, a representation in terms of
some finite set parameters, analogous to the $R^\pm$ of section 2.2, would
determine a quantization of those parameters.  For arbitrary $g$, it is not
yet clear exactly how to find such a representation, although for $g=2$
there has been some recent progress \cite{NR0}.

For the remainder of this section, we shall restrict our attention to the
relatively well-understood case of $g=1$, in order to make contact with
the torus moduli quantization of Section 4.1.  We can quantize the
classical algebra \rref{b7} as follows:
\begin{enumerate}
\item We replace the classical Poisson brackets  $\{\,,\,\}$ by commutators
$[\,,\,]$, with the rule
\beq
[x,y]= xy-yx =i \hbar \{x,y\} ;
\eeq
\item On the right hand side of \rref{b7}, we replace the product by
the symmetrized product,
\beq
xy \to \um (xy +yx) .
\eeq
\end{enumerate}
The resulting operator algebra is given by
\beq
\hat R_1^{\pm}\hat R_2^{\pm}e^{\pm i \theta}
  - \hat R_2^{\pm}\hat R_1^{\pm} e^{\mp i \theta}=
  \pm 2i\sin\theta\, \hat R_{12}^{\pm} \quad \hbox{\it and cyclical
  permutations}
\label{za}
\eeq
with
\beq
\tan\theta= { {i \sqrt k\hbar} \over {8\a}} .
\label{za1}
\eeq
Note that for $\Lambda\!<\!0$, $k=-1$, and $\theta$ is real, while for
$\Lambda\!>\!0$, $k=1$, and $\theta$ is pure imaginary.

The algebra \rref{za} is not a Lie algebra, but it is related to the
Lie algebra of the quantum group $\hbox{SU}(2)_q$ \cite{NRZ,su}, where
$q=\exp{4i\theta}$, and where the cyclically invariant $q$-Casimir is
the quantum analog of the cubic polynomial \rref{b9},
\beq
\hat F^{\pm}(\theta)
  = {\cos}^2\theta- e^{\pm 2i\theta} \left( (\hat R_1^\pm)^2+
(\hat R_{12}^\pm)^2\right) -e^{\mp 2i\theta} (\hat R_2^\pm)^2
 + 2e^{\pm i\theta}\cos\theta \hat R_1^\pm \hat R_2^\pm \hat R_{12}^\pm .
\eeq
The operator algebra \rref{za} can be represented by
\beq
\hat R_a^{\pm} = \sec\th \cosh{\hat r_a^{\pm} \over 2} \qquad (a=1,2,12) ,
\label{dc3}
\eeq
with
\beq
[\hat r_1^{\pm}, \hat r_2^{\pm}] = \pm 8i\th,
  \qquad[\hat r^+_a,\hat r^-_b] = 0 ,
\label{dc2}
\eeq
which differ (for $\Lambda$ small and negative) from the naive expectation
\beq
[\hat r_1^\pm, \hat r_2^\pm] = \mp {i\hbar\over\alpha} .
\label{dc1}
\eeq
by terms of order $\hbar^3$.

We must next try to implement the action of the modular group \rref{xa00}
on the operators $\hat R^\pm_a$.  The action that preserves the
commutators \rref{za} is (note the factor ordering)
\begin{eqnarray}
&S&:\hat R_1^{\pm}\rightarrow \hat R_2^{\pm},
  \quad \hat R_2^{\pm}\rightarrow \hat R_1^{\pm},
  \quad \hat R_{12}^{\pm}\rightarrow
        \hat R_1^{\pm} \hat R_2^{\pm} + \hat R_2^{\pm} \hat R_1^{\pm}
        - \hat R_{12}^{\pm}\nonumber\\
&T&:\hat R_1^{\pm}\rightarrow \hat R_{12}^{\pm},\quad
  \hat R_2^{\pm}\rightarrow \hat R_2^{\pm}, \quad
  \hat R_{12}^{\pm}\rightarrow \hat R_{12}^{\pm}\hat  R_2^{\pm}
        + \hat R_2^{\pm} \hat R_{12}^{\pm} - \hat R_1^{\pm} .
\label{za3}
\end{eqnarray}
The second of these can be generated by the unitary operators
\beq
G^{\pm} = \exp \bigl ( \pm  {i(\hat r_2^{\pm})^2 \over 16\theta} \bigr )
\eeq
as
$$
y \to G^{\pm} y (G^{\pm})^{-1}
$$
where $y$ is any function of the $\hat r_a^{\pm}$.

It is amusing to note that, from \rref{dc2}, the operators
$\exp\{n\hat r_1^{\pm}\}$ and $\exp\{\hat r_2^{\pm}\}$ commute when
$$\theta=\pi p/4n$$
with $n,p\!\in\!\IZ$ and $\Lambda\!<\!0$.  This occurs when the parameter
$q$ of the quantum group associated with the algebra \rref{za} is a root
of unity.  We see from \rref{za1} that there are $2n-1$ solutions $\alpha$
of this equation for for any given $n$.  By contrast, in the direct
quantization given by equation \rref{dc1}, this simplification of the
algebra would occur for an infinite number of values of $\alpha$.

\subsection{Quantizing the Space of Solutions}

A third method of quantization starts with the parameters $r_a^\pm$
of the classical solution \rref{c1}--\rref{c2}.  This approach can be
viewed as a version of covariant canonical quantization, i.e., ``quantizing
the space of classical solutions'' \cite{Ashcov,Crn}.  It has the obvious
disadvantage of requiring detailed knowledge of the classical solutions,
which are completely understood at present only for the simplest topology,
$\IR\!\times\!T^2$.  On the other hand, this approach to quantization
provides a natural bridge between the ADM and holonomy approaches
discussed above, and in particular allows us to define a natural set
of time-dependent physical operators in the latter theory.

Our starting point is now the set of Poisson brackets \rref{a1}.  The
natural guess is that these should simply become the commutators
\rref{dc1}. This leads to a legitimate quantum theory, but we know from
the previous section that the commutators \rref{za} of traces of
holonomies will not be reproduced. To obtain these traces, we must instead
impose the commutators \rref{dc2}. With these definitions, the results
of the previous section are all preserved.  In particular, it is not
hard to show that the quantum modular group action \rref{za3} is induced
by the transformations \rref{xa00} of the $\hat r^\pm_a$.

We can now make the connection with the ADM quantization of section 4.1.
The basic idea is to treat wave functions $\psi(r_a)$ as Heisenberg picture
states and to define suitable time-dependent operators acting on these
states.  Now, the {\em classical} modulus and momentum on a surface $K\!=\!
\tau$ have already been determined in terms of the $r^\pm_a$, and are given
by equations \rref{c3}--\rref{c4}.  Carrying these definitions over to the
quantum theory, we obtain a family of operators $\hat m(\tau)$ and
$\hat p(\tau)$, whose eigenvalues may be interpreted as the ADM modulus
and momentum in the York time slicing.  Similarly, the operator analog
of \rref{a2} may be interpreted as a Hamiltonian generating the evolution
of $\hat m$ and $\hat p$.  Indeed, if we keep the orderings of
\rref{c3}--\rref{c4} and \rref{a2}, defining
\begin{eqnarray}
\hat m &=& \left(\hat r_1^-e^{it/\a} + \hat r_1^+e^{-{it/\a}}\right)
 \left(\hat r_2^-e^{it/\a} + \hat r_2^+e^{-{it/\a}}\right)^{\lower2pt%
 \hbox{$\scriptstyle -1$}} \nonumber\\
\hat p &=& -{i\a\over 2\sin{2t\over \a}}\left(\hat r_2^+e^{it/\a}
 + \hat r_2^-e^{-{it/\a}}\right)^{\lower2pt\hbox{$\scriptstyle 2$}}\\
\hat H &=& {\a^2\over 4}\sin{2t\over \a}\,
 (\hat r_1^-\hat r_2^+ - \hat r_1^+\hat r_2^-), \nonumber
\label{dc4}
\end{eqnarray}
it follows from the commutators \rref{dc2} that
\begin{eqnarray}
&[&\hat m^\dagger,\hat p]=[\hat m,\hat p^\dagger]= 16i\a\th, \qquad
  [\hat m,\hat p] = [\hat m^\dagger,\hat p^\dagger] = 0\\
&[&\hat p, \hat H^\prime] = -8i\a\th {d\hat p\over dt},
  \qquad  [\hat m, \hat H^\prime] = -8i\a\th {d\hat m\over dt} ,
\label{dc5}
\end{eqnarray}
which differ from the corresponding equations in ADM quantization by terms
of order $O(\hbar^3)$, small when $|\Lambda|\!=\!1/\alpha^2$ is small.
These results depend on operator ordering in $\hat m$, $\hat p$, and
$\hat H$, of course, but the orderings of \rref{dc4} are a bit less
arbitrary than they might seem: they were chosen to ensure that the
modular transformations \rref{xa00} of the $\hat r^\pm_a$ induce the
correct transformations \rref{bb15} of $\hat m$ without any $O(\hbar)$
corrections.

Note that \rref{dc3} can be inverted to give the operators $\hat r^\pm_a$
in terms of the traces $\hat R^\pm_a$.  Equation \rref{dc4} can therefore
be viewed as a definition of modulus and momentum operators in the
holonomy algebra quantization.  These operators are, of course, quite
complicated---they involve logarithms of the traces $\hat R$---but given
a representation of the holonomy algebra, they provide the first known
instance of physical observables with clear geometric interpretations.

To further investigate the connection to ADM quantization, we can examine
the properties of wave functions that are eigenfunctions of $\hat
m(r^\pm_a,\tau)$ and its adjoint---that is, we can transform to a
``Schr\"odinger picture.''  As in the $\Lambda=0$ case \cite{dirac},
Ezawa has shown that these wave functions transform as Maass forms
of weight $1/2$, corresponding to an ordering \rref{da5} of the ADM
Hamiltonian with $n=1/2$ \cite{Ezawa1}.  Also as in the $\Lambda=0$ case
\cite{ordering}, however, this Hamiltonian can be changed by reordering the
operators $\hat m^\dagger(r^\pm_a,\tau)$ and $\hat p^\dagger(r^\pm_a,\tau)$,
or equivalently by redefining the inner product.

We can now return to the question of the choice of time slicing raised
at the end of section 4.1.  In the holonomy quantization of section 4.2
or the approach of this section, no choice of a time coordinate is ever
made.  A particular time slicing is instead reflected in a choice of
``time''-dependent operators $\hat m(r^\pm_a,\tau)$ and $\hat p(r^\pm_a,
\tau)$ that describe the geometry of the chosen slice.  Other choices of
classical time coordinate would presumably lead to other operators, which
would be used to answer genuinely different physical questions.  In some
sense, we have thus succeeded in evading the ``problem of time'' in
quantum gravity.

\section{Conclusion}

In most quantum field theories, it is fairly clear from the start
what the ``right'' variables to quantize are.  Moreover, we have
general theorems that guarantee that local field redefinitions will
not change the $S$ matrix.  Consequently, we are not used to worrying
about how to determine the ``right'' quantization of, say,
electrodynamics.

Quantum gravity is different.  Here, the physical observables are
necessarily nonlocal, and there is no reason to believe that
quantizations based on different variables should be equivalent.
In (3+1)-dimensional gravity, of course, the question is rather
premature, since we do not yet have even one complete, consistent
quantization.  In 2+1 dimensions, though, the problem becomes
unavoidable.

It might be hoped, however, that this problem can be turned to our
advantage.  The various approaches to quantizing (2+1)-dimensional
gravity have different strengths, and if their relationship can be
understood clearly, we might be able to combine these strengths.  In
ADM quantization, for instance, the fundamental variables---the moduli
and momenta $(m,p)$---have simple geometric interpretations.  In the
quantization of the traces $R^\pm$, the physical meaning of the
observables is much less clear, but the algebraic structures can
be directly generalized to arbitrary spatial topologies.  A primary goal
of this paper has been to demonstrate the relationships between these
approaches, thus allowing us to introduce clearly defined physical
observables into the algebraic structure of holonomy quantization.

As we have seen, this goal can be achieved for spacetimes of the
form $\IR\!\times\!T^2$.  Equations \rref{dc3} and \rref{dc4}
give explicit ``time''-dependent operators in the holonomy formalism
that represent the moduli and momenta on surfaces of constant York
time.  These results depend on our knowledge of the exact solutions
of the equations of motion, but it may be possible to extend them
at least to genus $2$: reference \cite{NR0} has developed the description
of the quantum algebra of traces of holonomies, while the hyperelliptic
nature of genus $2$ surfaces is likely to simplify the ADM analysis.

We have also seen hints of a solution of the ``problem of time'' in
quantum gravity.  In ADM quantization, one must choose a classical
time-slicing, and it is by no means clear that different choices will
lead to equivalent theories.  In quantization of the holonomy algebra,
on the other hand, no such choice need be made; different choices of time
show up only as different families of operators describing the spatial
geometry of the corresponding slices.  The definition of such operators
is difficult, of course, and it would be very useful to find a perturbative
approach that did not require complete knowledge of the classical
solutions.  But in principle, we have found a way to implement Rovelli's
approach to ``evolving constants of motion'' \cite{Rov} in a theory of
quantum gravity.

\vspace{1.5ex}
\begin{flushleft}
\large\bf Acknowledgements
\end{flushleft}

This work was supported in part by National Science Foundation grant
PHY-93-57203, Department of Energy grant DE-FG03-91ER40674, and INFN
Iniziativa Specifica TO10\ (FI2).

\section{Appendix}
\setcounter{footnote}{0}

In this appendix, we briefly describe some conventions of this paper.
The $\gamma$-matrices are defined by
\begin{eqnarray}
\ga_{A}\ga_{B}+\ga_{B}\ga_{A}&=& 2g_{AB}\nonumber\\
 \ga_{AB}&=&\um [\ga_{A},\ga_{B}]
\end{eqnarray}
and $\ga^{\ }_{5} =k \ga_{_0}\ga^{\ }_{1}\ga^{\ }_{2}\ga^{\ }_{3}$
satisfies:
\beq
\ga_{5}^{2}=-k,\ -\um \ep^{ABCD}\ga_{A}\ga_{B}=\ga^{\ }_{5}\ga^{C}
\ga^{D} .
\eeq
In terms of the Pauli matrices, our representation is
$$
\gamma_0=i\sigma_2\otimes \sigma_3,\
\gamma_1= \sigma_3\otimes \sigma_3,\
\gamma_2=-1\otimes \sigma_1,\
\gamma_3=-\sqrt{k} \sigma_1\otimes \sigma_3,\
\gamma_5=i\sqrt{k}\, 1\otimes \sigma_2.
$$

The projectors
\beq
P_{\pm}=\pm(\ga^{\ }_{5}\pm i\sqrt{k} 1)/(2i\sqrt{k})
\eeq
satisfy $P_{\pm}^{2}=P_{\pm}$, $P_{+}P_{-}=P_{-}P_{+}=0$. In this
representation,
\beq
P_{\pm}=1\otimes p_{\pm},\quad p_{\pm}=\um (1\pm\sigma_2) .
\eeq
It follows that
\beq
\gamma^{AB}\otimes \gamma_{AB}
  =-4(\sigma^{i}\otimes p_{+})\otimes(\sigma^{i}\otimes p_{+})
   -4(\sigma^{i}\otimes p_{-})\otimes(\sigma^{i}\otimes p_{-})
\eeq
where
\beq
{(\sigma^i)_{\alpha}}^{\beta} {(\sigma^i)_{\gamma}}^{\tau}
=-\da_{\alpha}^{\beta}\da_{\gamma}^{\tau}
+2\da_{\alpha}^{\tau}\da^{\beta}_{\gamma}.
\eeq

\end{document}